\documentclass[3p]{elsarticle}
\usepackage{hyperref}
\usepackage{graphicx}
\usepackage{enumitem}
\usepackage{multirow}
\usepackage{makecell}
\usepackage{url}
\usepackage{subcaption}
\usepackage{color, colortbl}
\RequirePackage{amssymb,amsmath,bm}
\RequirePackage{textcomp}
\RequirePackage{booktabs}
\usepackage{marginnote}
\usepackage{inputenc}

\usepackage[table,xcdraw]{xcolor}
\usepackage[normalem]{ulem}
\useunder{\uline}{\ul}{}
\definecolor{Gray}{gray}{0.9}
\newcolumntype{g}{>{\columncolor{Gray}}c}
           
\makeatletter
\def\ps@pprintTitle{%
 \let\@oddhead\@empty
 \let\@evenhead\@empty
 \def\@oddfoot{%
    \footnotesize 
    \parbox{\textwidth}{%
        \textit{Paper published in Computer Speech \& Language.\\
	        The version of record is available at} \url{https://doi.org/10.1016/j.csl.2024.101685}
    }\hfill
 }
 \let\@evenfoot\@oddfoot}
\makeatother

\makeatletter
\def\printFirstPageNotes{%
  \iflongmktitle
    \let\columnwidth=\textwidth
  \fi
\ifdoubleblind
\else
  \ifx\@tnotes\@empty\else\@tnotes\fi
  \ifx\@nonumnotes\@empty\else\@nonumnotes\fi
  \ifx\@cornotes\@empty\else\@cornotes\fi
  \ifx\@elseads\@empty\relax\else
   \let\thefootnote\relax
   \footnotetext{\ifnum\theead=1\relax
      \textit{Email address:\space}\else
      \textit{Email addresses:\space}\fi
     \@elseads}\fi
  \ifx\@elsuads\@empty\relax\else
   \let\thefootnote\relax
   \footnotetext{\textit{URL:\space}%
     \@elsuads}\fi
\fi
  \ifx\@fnotes\@empty\else\@fnotes\fi
  \iflongmktitle\if@twocolumn
   \let\columnwidth=\Columnwidth\fi\fi
}
\makeatother

\journal{}

\hypersetup{
    colorlinks=true,
    linkcolor=blue,
    filecolor=magenta,      
    urlcolor=cyan
}

\biboptions{authoryear}

\newcommand{\revision}[1]{#1}

\begin{document}

\begin{frontmatter}

\title{Objective and subjective evaluation of speech enhancement methods in the UDASE task of the 7th CHiME challenge}

\author[1]{Simon Leglaive\corref{cor1}}
\ead{simon.leglaive@centralesupelec.fr}
\author[2]{Matthieu Fraticelli}
\author[3]{Hend ElGhazaly}
\author[4]{Léonie Borne}
\author[5]{Mostafa Sadeghi}
\author[6]{\mbox{Scott Wisdom}}
\author[4]{Manuel Pariente}
\author[6]{John R. Hershey}
\author[2]{Daniel Pressnitzer}
\author[3]{Jon P. Barker}
%\ead{}
\address[1]{CentraleSup\'elec, IETR UMR CNRS 6164, France}
\address[2]{\'Ecole Normale Sup\'erieure, PSL University, CNRS, France}
\address[3]{Department of Computer Science, University of Sheffield, UK}
\address[4]{Pulse Audition, France}
\address[5]{Inria, France}
\address[6]{Google, USA}
\cortext[cor1]{Corresponding author}
%\tnotetext[t1]{}
\begin{abstract}
Supervised models for speech enhancement are trained using artificially generated mixtures of clean speech and noise signals. However, the synthetic training conditions may not accurately reflect real-world conditions encountered during testing. This discrepancy can result in poor performance when the test domain significantly differs from the synthetic training domain. To tackle this issue, the UDASE task of the 7th CHiME challenge aimed to leverage real-world noisy speech recordings from the test domain for unsupervised domain adaptation of speech enhancement models. Specifically, this test domain corresponds to the CHiME-5 dataset, characterized by real multi-speaker and conversational speech recordings made in noisy and reverberant domestic environments, for which ground-truth clean speech signals are not available. In this paper, we present the objective and subjective evaluations of the systems that were submitted to the CHiME-7 UDASE task, and we provide an analysis of the results. This analysis reveals a limited correlation between subjective ratings and several supervised nonintrusive performance metrics recently proposed for speech enhancement. Conversely, the results suggest that more traditional intrusive objective metrics can be used for in-domain performance evaluation using the reverberant LibriCHiME-5 dataset developed for the challenge. The subjective evaluation indicates that all systems successfully reduced the background noise, but always at the expense of increased distortion. Out of the four speech enhancement methods evaluated subjectively, only one demonstrated an improvement in overall quality compared to the unprocessed noisy speech, highlighting the difficulty of the task. The tools and audio material created for the CHiME-7 UDASE task are shared with the community.
\end{abstract}


\end{frontmatter}

% \linenumbers

\section{Introduction}

The speech enhancement task consists of improving the quality and intelligibility of a degraded speech signal recording. One approach to achieve this is through noise suppression algorithms, which aim to estimate the clean speech signal by removing the additive background noise in the recording. 
Over the past 50 years, traditional signal-processing-based speech enhancement algorithms \citep{boll1979suppression, lim1979enhancement, ephraim1984speech, martin2005speech, loizou2013speech} have been progressively outperformed by data-driven approaches using hidden Markov models (HMMs) \citep{sameti1998hmm, ephraim1992bayesian,sameti1998hmm}, codebook-based approaches \citep{srinivasan2005codebook}, non-negative matrix factorization (NMF) \citep{mohammadiha2013supervised}, and more recently deep learning \citep{wang2017supervised}. 

There has been great progress in speech enhancement in recent years thanks to deep learning models trained in a \emph{supervised} manner. The conventional approach in supervised speech enhancement involves three main ingredients:
\begin{enumerate}[]
    \item A \emph{model}, which provides a prediction of the clean speech signal given the noisy recording. Current state-of-the-art methods rely on deep neural networks.
    \item A \emph{metric}, which measures the discrepancy between the clean speech estimate and the ground-truth signal. During training, the metric corresponds to a differentiable loss function, which is minimized to estimate the model parameters. At test time, the metric (which can differ from the training loss function) is used to evaluate the performance of the trained model.
    \item A \emph{labeled dataset}, which consists of noisy speech signals paired with their corresponding clean reference signals. During training, these clean reference signals serve as the model's targets, and at test time, they are used to compute intrusive performance metrics.
\end{enumerate}
Unfortunately, it is very difficult, if not impossible, to acquire labeled noisy speech signals in real-world conditions due to cross-talk between microphones. Datasets for supervised learning and performance evaluation with intrusive metrics have to be generated artificially, by creating synthetic mixtures of isolated speech and noise signals. 

A large research effort has been put recently on the three main ingredients of supervised speech enhancement (the model, the metric, and the labeled dataset), and it is undeniable that this effort has led to unprecedented results, e.g., \citet{weninger2015speech, fu2017raw, pascual2017segan, choi2018phase, zhao2018convolutional, fu2019metricgan, defossez2020real, cosentino2020librimix, hao2021fullsubnet, pandey2021dense, fu2021metricgan+, richter2023speech}. However, the fully supervised learning paradigm also has limitations when applied to speech enhancement. First, creating a synthetic dataset of realistic noisy speech mixtures is not easy and requires important engineering efforts. Second, supervised speech enhancement is effective as long as the acoustic recording conditions at test time are covered by the synthetic training data. This is a constraint that is hard to satisfy due to the variability of the acoustic recording conditions, in terms of noise type, signal-to-noise ratio, recording equipment, speaker-to-microphone distance and orientation, reverberation, etc. Supervised speech enhancement performance can thus decrease significantly in case of mismatch between the training and testing conditions \citep{pandey2020cross,bie2022unsupervised,richter2023speech,gonzalez2023assessing}. Finally, when the test domain differs from the synthetic training domain, supervised learning necessitates rebuilding the synthetic training dataset and retraining the model, which is time-consuming and computationally intensive. A more effective approach would be to automatically adapt models to real, unlabeled noisy speech recordings, eliminating the need for ground-truth clean speech signals.

The ability to adapt to unknown adverse acoustic conditions while perceiving speech is a fundamental property of the human auditory system \citep{bent2009perceptual, brandewie2010prior,cooke2022time}, which cannot be reproduced in machine listening systems using a fully supervised approach. Adaptation of speech enhancement systems using real unlabeled noisy speech data is the main challenge that the CHiME-7 UDASE\footnote{Unsupervised Domain Adaptation for conversational Speech Enhancement.} task tried to address \citep{leglaive2023chime}. Previous challenges for single-channel speech enhancement, such as the deep noise suppression (DNS) challenges \citep{reddy2020interspeech, reddy2021icassp, reddy2021interspeech, dubey2022icassp}, focused on a supervised setup with a training set consisting of a large amount of labeled synthetic data intended to cover diverse conditions. The CHiME-7 UDASE task was intended to study a different situation, targeting single-channel speech enhancement in a specific domain for which no well-matched labeled data are available for training. 

In \citet{leglaive2023chime}, we introduced the task and described the data and the baseline system. In the present paper, we extend the description of the CHiME-7 UDASE task by introducing the speech enhancement methods that were submitted to the challenge, describing their objective and subjective evaluation, and providing an analysis of the results. Along with this paper, we release the JavaScript experimental platform we developed for the listening test, following the \citet{recommendation2003subjective} methodology. We also release the audio files that were submitted to the challenge along with the corresponding human listening scores (in addition to various objective evaluation scores), which could serve as a voice quality dataset for speech enhancement research \citep{leglaive_2024_10418311}. To the best of our knowledge, this is the first dataset of \citet{recommendation2003subjective} subjective evaluation results made publicly available.

The paper is organized as follows. The task and datasets are presented in Section~\ref{sec:task_and_dataset}. The objective and subjective evaluation protocols are described in Sections~\ref{sec:objective_eval} and \ref{sec:listening_test}, respectively. Section~\ref{sec:SE_methods} introduces the speech enhancement methods that participated in the CHiME-7 UDASE task. The results of the subjective and objective evaluations are presented and analyzed in Section~\ref{sec:results} before concluding in Section~\ref{sec:conclusion}.

\section{Task and datasets}
\label{sec:task_and_dataset}

\subsection{Task}

The CHiME-7 UDASE task focuses on single-channel speech enhancement in a specific target domain for which no well-matched labeled training data are available. It consists of using unlabeled data in the target domain to adapt supervised speech enhancement models trained on synthetic labeled data in a source domain. The target domain corresponds to the real conversational speech recordings of the CHiME-5 dataset \citep{barker2018fifth}. These recordings were made during dinner parties in real homes, with multiple speakers in noisy and reverberant environments. Given a mixture of one or more reverberant speakers and additive background noise, the objective of this task is to estimate the clean, potentially multi-speaker, reverberant speech, removing the additive background noise. The task is motivated by an assistive listening use case, in which a speech enhancement algorithm can help any individual to better engage in a conversation, by improving the overall speech quality and intelligibility within the ambient noise.

\subsection{Datasets}

The CHiME-7 UDASE task is built upon three datasets:
\begin{enumerate}
    \item The CHiME-5 dataset \citep{barker2018fifth} for the in-domain unlabeled data, which are used for training, development, and evaluation;
    \item The LibriMix dataset \citep{cosentino2020librimix} for the out-of-domain labeled data, which are used for training and development only;
    \item A new reverberant LibriCHiME-5 dataset, which was created to provide ``close to in-domain" labeled data for development and evaluation only.
\end{enumerate}
All three datasets include mixtures of reverberant speech and noise, with up to three overlapping speakers. All signals are sampled at 16 kHz. The datasets are presented in the rest of this section, and additional information can be found in \citet{leglaive2023chime}.

\subsection{CHiME-5 in-domain unlabeled data} 
\label{subsec:chime-5}

The CHiME-5 dataset \citep{barker2018fifth} consists of recordings made during twenty real dinner parties (or sessions) of between two and three hours. Each dinner party involved four participants wearing binaural microphones and took place in a different home, with three recording locations per home (kitchen, dining room, living room). The CHiME-5 recordings include natural conversations between multiple speakers in reverberant and noisy environments, and they are fully transcribed. Using the CHiME-5 transcription files, we estimated that 22\% of the audio recordings contain only noise, 51\% contain one single active speaker, and 20\%, 5\%, and 2\% contain two, three, and four overlapping speakers, respectively.\footnote{These numbers were obtained without considering any constraint on the overlap duration and using a labeling precision of 0.01 second.} The CHiME-7 UDASE task only uses the binaural recordings (Kinect recordings are also available), from which single-channel audio segments with up to 3 overlapping speakers and background noise were extracted. 

The dataset contains training, development, and evaluation sets, with different speakers in each set. The training set consists of the raw single-channel audio segments extracted from the binaural recordings when the participant wearing the microphone does not speak. It is intended to be used for unsupervised adaptation on in-domain unlabeled noisy speech data.

No ground-truth clean speech signals are available for the CHiME-5 noisy speech recordings. This is a major difficulty because for developing and evaluating a speech enhancement system one needs to compute objective performance metrics, which is usually achieved using ground-truth signals. To circumvent this difficulty, the transcription files were used to segment the CHiME-5 recordings in short audio segments labeled with the maximum number of simultaneously active speakers (0, 1, 2 or 3).\footnote{This only corresponds to a maximum value, i.e., through the duration of a segment the number of simultaneously-active speakers can vary between 0 and the maximum value. Moreover, a segment might contain more speakers than the labeled maximum number of simultaneously active speakers. For instance, a segment labeled as a single-speaker might contain two active speakers who do not speak simultaneously.} This segmentation was done only for the development and evaluation sets, which simulates the reasonable scenario where we can afford to manually annotate a small amount of data with speaker count labels for development and evaluation, but this procedure cannot be easily done for a large training set. The noise-only segments were subsequently used to create the synthetic labeled noisy speech mixtures of the reverberant LibriCHiME-5 dataset (see Section~\ref{subsec:librichime-5}). The single-speaker segments can be used to compute nonintrusive (reference-free) objective performance metrics, meaning they do not require ground-truth clean speech signals. The dataset also contains an evaluation subset intended to be used for the listening test. It consists of test audio samples extracted by looking for segments of 4 to 5 seconds with at least 3 seconds of speech and 0.25 seconds without speech at the beginning and the end. Additional constraints were taken into account to ensure a balanced subset in terms of the speaker's gender, recording location, and session.

The segmented CHiME-5 dataset used for the \mbox{CHiME-7} UDASE task is summarized in Table~\ref{tab:chime-5}. Additional details about the segmentation of the CHiME-5 original recordings are provided in \citet{leglaive2023chime}, and the scripts to generate the dataset are available online.\footnote{\url{https://github.com/UDASE-CHiME2023/CHiME-5} (last accessed: 2024-02-02)}

\begin{table}[t]
\centering
\resizebox{.5\linewidth}{!}{ 
\begin{tabular}{ccccc}
\cmidrule[.8pt]{1-5}
& & \multicolumn{2}{c}{\makecell{Sample length (s)}}          & Total duration \\
Subset & \# samples & Mean & SD &     (HH:MM:SS)                                       \\
\cmidrule(lr){1-1} \cmidrule(lr){2-2} \cmidrule(lr){3-3} \cmidrule(lr){4-4} \cmidrule(lr){5-5}
\texttt{train}           & 27 517                                                  & 10.91                    & 14.10                   & 83:22:29                                                       \\ \cmidrule(lr){1-5}
\texttt{dev/0}           & 912                                                    & 6.50                     & 4.10                    & 1:38:49                                                        \\
\texttt{dev/1}           & 5 719                                                   & 5.89                     & 3.49                    & 9:21:53                                                        \\
\texttt{dev/2}           & 3 835                                                   & 5.23                     & 2.43                    & 5:34:33                                                        \\
\texttt{dev/3}           & 667                                                    & 4.61                     & 1.84                    & 0:51:14                                                        \\ \cmidrule(lr){1-5}
\texttt{eval/0}          & 977                                                    & 5.73                     & 3.35                    & 1:33:19                                                        \\
\texttt{eval/1}          & 3 013                                                   & 5.54                     & 2.94                    & 4:35:05                                                        \\
\texttt{eval/2}          & 1 552                                                   & 4.88                     & 2.04                    & 2:06:07                                                        \\
\texttt{eval/3}          & 233                                                    & 4.21                     & 1.17                    & 0:16:21                                                        \\
\texttt{eval/LT} 		& 241                                                    & 4.72                     & 0.34                    & 0:18:58   \\ \cmidrule[.8pt]{1-5}
\end{tabular}
}
\caption{Segmented CHiME-5 dataset. Dev and eval subsets are labeled with the maximum number of simultaneously active speakers (0, 1, 2, 3). \texttt{eval/LT} corresponds to the evaluation subset for the listening test.}
\label{tab:chime-5}
\end{table}

\subsection{LibriMix out-of-domain labeled data}
\label{subsec:librimix}

The CHiME-7 UDASE task employs the LibriMix dataset \citep{cosentino2020librimix} for supervised learning on out-of-domain data. LibriMix was originally developed for speech separation in noisy environments, and it is derived from LibriSpeech ``clean'' utterances \citep{panayotov2015librispeech} and WHAM! noises \citep{wichern2019wham}. The Libri2Mix and Libri3Mix versions of the dataset contain noisy speech mixtures with 2 and 3 overlapping speakers, respectively. A single-speaker version of LibriMix (Libri1Mix) can be obtained by simply discarding one of the two speakers in Libri2Mix mixtures. A complete description of LibriMix is provided by \cite{cosentino2020librimix}.

\subsection{Reverberant LibriCHiME-5 close-to-in-domain labeled data}
\label{subsec:librichime-5}

\begin{table}[]
\centering
\resizebox{.5\linewidth}{!}{ 
\begin{tabular}{cccccc}
\cmidrule[.8pt]{1-6}
& & & & \multicolumn{2}{c}{RT60 (s)} \\
Subset & Home & Room & \# RIRs & Mean & SD \\
\cmidrule(lr){1-1} \cmidrule(lr){2-2} \cmidrule(lr){3-3} \cmidrule(lr){4-4} \cmidrule(lr){5-5}  \cmidrule(lr){6-6}
\multirow{5}{*}{\texttt{dev}} & 2 & Living room & 160 & 0.582 & 0.023 \\
 & 2 & Kitchen & 120 & 0.500 & 0.022 \\
 & 2 & Bedroom & 136 & 0.471 & 0.056 \\
 & 3 & Living room & 136 & 0.437 & 0.026 \\
 & 3 & Bedroom & 120 & 0.431 & 0.036 \\
\cmidrule{1-6}
\multirow{4}{*}{\texttt{eval}} & 3 & Kitchen & 128 & 0.451 & 0.037 \\
 & 4 & Living room & 144 & 0.460 & 0.034 \\
 & 4 & Kitchen & 128 & 0.375 & 0.035 \\
 & 4 & Bedroom & 128 & 0.387 & 0.022 \\
\cmidrule[.8pt]{1-6}
\end{tabular}
}
\caption{Estimated RT60s (in seconds) of the rooms in the VoiceHome corpus \citep{bertin2016french} that were used to create the reverberant LibriCHiME-5 data (\texttt{dev} and \texttt{eval} subsets). The column `\# RIRs' gives the number of single-channel RIRs used to compute the mean and standard deviation (SD) of the RT60.}
\label{tab:RT60}
\end{table}

We created the reverberant LibriCHiME-5 dataset to provide ``close-to-in-domain'' labeled data for computing standard intrusive objective performance metrics used in speech enhancement, such as the scale-invariant signal-to-distortion ratio (SI-SDR) \citep{leroux2019sdr}. This dataset consists of synthetic mixtures of reverberant speech and noise, with up to three simultaneously active speakers, labeled with the clean reference speech signals. In-domain noise signals were extracted from the CHiME-5 recordings using the ground-truth transcriptions (see Section~\ref{subsec:chime-5}), and clean speech utterances were taken from the LibriSpeech dataset \citep{panayotov2015librispeech} and convolved with room impulse responses (RIRs) from the VoiceHome corpus \citep{bertin2016french}. \revision{These RIRs were recorded in 12 different rooms of 3 real homes, with 4 rooms per home: living room (room 1), kitchen (room 2), bedroom (room 3), and bathroom (room~4). Bathrooms were excluded for the reverberant LibriCHiME-5 dataset. In each room, RIRs were recorded for 2 different positions and geometries of an 8-channel microphone array and 7 to 9 different positions of the loudspeaker.
Table~\ref{tab:RT60} indicates the estimated reverberation times (RT60s) of the rooms in the VoiceHome corpus whose RIRs were used to create the reverberant LibriCHiME-5 dataset. The RT60 is defined as the time it takes for the sound energy to decrease by 60~dB after the extinction of the source. For each room in each home, we estimated the RT60 on each single-channel RIR by fitting a straight line on Schroeder’s integrated energy decay curve using linear least-squares regression \citep{schroeder1965new}. We then computed the mean RT60 and the standard deviation for each room.
}

\revision{The process to create the synthetic reverberant LibriCHiME-5 dataset was the following. For each mixture in the dataset, we randomly chose the maximum number $n \in \{1,2,3\}$ of simultaneously active speakers in the mixture, with $p(n=i) = 0.60, 0.35, 0.05$ for $i=1,2,3$, respectively, which is consistent with the distribution of the segmented CHiME-5 dataset. Each mixture's speakers were randomly sampled from the list of LibriSpeech speakers with equal probability to be a male or a female. We used the VoiceHome corpus to simulate the acoustic recording environment in the reverberant LibriCHiME-5 dataset. For each mixture, we randomly and successively sampled a home, a room, an array position/geometry, $n$ speaker positions without replacement, and a channel of the microphone array, which gave a set of RIRs. LibriSpeech utterances were convolved with the selected RIRs to obtain the reverberant speech utterances. These were then mixed following speech activity patterns extracted from the CHiME-5 transcription files (diarization labels) to simulate a natural conversation between multiple speakers. This synthetic mixing of multi-speaker speech involved selecting and trimming LibriSpeech utterances in order to make them fit in the CHiME-5 activity patterns. 
% In particular, the end (resp. beginning) of a utterance was used when it had to be located at the beginning (resp. at the end or in the middle) of a mixture. 
The multi-speaker reverberant speech and noise mixtures were finally created such that the per-speaker signal-to-noise ratio (SNR) was distributed as a Gaussian with a mean of 5~dB and a standard deviation of 7~dB to match the SNR distribution of the CHiME-5 dataset as estimated by Brouhaha \citep{lavechin2022brouhaha}. This was achieved by first sampling a global per-mixture SNR $x \sim \mathcal{N}(5, \sigma_1^2)$ and then sampling a local per-speaker SNR $y_n \sim \mathcal{N}(x, \sigma_2^2)$, with $\sigma_1 = 6.7082$ and $\sigma_2 = 2$ ($\sqrt{\sigma_1^2 + \sigma_2^2} \approx 7$~dB). The value of $\sigma_2$ was chosen such that the loudness difference between multiple speakers remained moderate; this was again to simulate a conversation. A detailed description of the process implemented to create the reverberant LibriCHiME-5 mixtures (metadata and audio files) and the corresponding Python code are available online.}\footnote{\url{https://github.com/UDASE-CHiME2023/reverberant-LibriCHiME-5} (last accessed: 2024-04-03).}

Overall, in the reverberant LibriCHiME-5 dataset, the speech utterances were convolved with RIRs measured in real homes, the noise signals were extracted from the in-domain CHiME-5 recordings, the SNR was chosen to approximately match that of the target domain, and the speech utterances were mixed to simulate a conversation using the CHiME-5 transcription. It is therefore hoped that the performance on the reverberant {LibriCHiME-5} dataset corresponds to an estimate of the performance on the CHiME-5 dataset, which will be confirmed by the analysis of the evaluation results. 
The reverberant LibriCHiME-5 dataset is summarized in Table~\ref{tab:librichime-5}.

\begin{table}[t]
\centering
\resizebox{.5\linewidth}{!}{ 
\begin{tabular}{ccccc}
\cmidrule[.8pt]{1-5}
& & \multicolumn{2}{c}{\makecell{Sample length (s)}}          & Total duration \\
Subset & \# samples & Mean & SD &     (HH:MM:SS)                                        \\
\cmidrule(lr){1-1} \cmidrule(lr){2-2} \cmidrule(lr){3-3} \cmidrule(lr){4-4} \cmidrule(lr){5-5}
\texttt{dev/1}  & 1 187              & 7.14               & 4.67               & 2:21:09        \\
\texttt{dev/2}  & 565               & 5.37               & 2.24               & 0:50:31        \\
\texttt{dev/3}  & 65                & 4.81               & 1.66               & 0:05:12        \\ \cmidrule(lr){1-5}
\texttt{eval/1} & 1 394             & 6.25               & 3.75               & 2:25:17        \\
\texttt{eval/2} & 494               & 4.44               & 1.34               & 0:36:35        \\
\texttt{eval/3} & 64                & 4.21               & 1.07               & 0:04:29                                                       \\ \cmidrule[.8pt]{1-5}
\end{tabular}
}
\caption{Reverberant LibriCHiME-5 dataset. The subsets are labeled with the maximum number of simultaneously active speakers (0, 1, 2, 3).}
\label{tab:librichime-5}
\end{table}

\section{Objective evaluation}
\label{sec:objective_eval}

We consider several intrusive metrics for the objective evaluation of the systems that participated in the CHiME-7 UDASE task:
\begin{itemize}[]
    \item The SI-SDR (in dB) is a ubiquitous measure of audio source separation quality \citep{leroux2019sdr}, used in particular for denoising. It corresponds to an adaptation of the standard SNR that makes it invariant to an arbitrary scaling of the clean speech estimate.
    \revision{The SI-SDR is defined by:
    \begin{equation}
    \text{SI-SDR}(\mathbf{s}, \hat{\mathbf{s}}) = 10 \log_{10}\left(\frac{\parallel \beta \mathbf{s} \parallel^2}{\parallel \hat{\mathbf{s}} - \beta \mathbf{s} \parallel^2} \right), \qquad \beta = \frac{\left<\hat{\mathbf{s}}, \mathbf{s} \right>}{\parallel \mathbf{s}\parallel^2},
    \end{equation}
    where $\mathbf{s}, \hat{\mathbf{s}} \in \mathbb{R}^T$ denote respectively the ground-truth and estimated speech signals with $T$ samples, $\left<\hat{\mathbf{s}}, \mathbf{s} \right> = \hat{\mathbf{s}}^\top \mathbf{s}$, and $\parallel \mathbf{s}\parallel^2 = \left<\mathbf{s}, \mathbf{s} \right>$.
    }
    We use a straightforward Python implementation of this metric.
    \item The wideband perceptual evaluation of speech quality (PESQ) measure was developed to provide an estimate of the speech quality as it would be perceived by humans \citep{rix2001perceptual}. We use the \texttt{pesq} Python package \citep{miao_wang_2022_6549559}, which provides a MOS-LQO (Mean Opinion Score - Listening Quality Objective) score between 1.04 and 4.64 following the \citet{recommendation2007wideband} recommendation.
   \item The short-time intelligibility index (STOI, between 0 and 1) is a measure highly correlated with the intelligibility of degraded speech signals \citep{taal2010short,taal2011algorithm}. We use the \texttt{pystoi} Python package \citep{pariente_pystoi}.
\end{itemize}
The SI-SDR, PESQ, and STOI measures are very standard in speech enhancement, but as intrusive metrics, they require ground-truth clean speech, which is not available for the in-domain CHiME-5 recordings. Therefore, we also consider several nonintrusive learning-based metrics for objective performance evaluation:
\begin{itemize}[]
    \item DNSMOS P.835 (hereinafter DNSMOS) provides an estimate of the three scores of the \citet{recommendation2003subjective} subjective evaluation methodology, which assesses the speech signal quality (SIG), the background intrusiveness (BAK), and the overall quality (OVRL) \citep{reddy2022dnsmos}. DNSMOS was trained using the human listening scores crowdsourced in the context of the DNS Challenge 2021 \citep{reddy2021interspeech}, which were used as supervised training targets. We use a functional implementation of the DNSMOS authors' code \citep{reddy2022dnsmos}.
    \item TorchAudio-Squim (TAS) provides nonintrusive estimates of the SI-SDR, wideband PESQ, and STOI measures \citep{kumar2023torchaudio}, denoted by TAS SI-SDR, TAS PESQ, and TAS STOI. These metrics were trained in a supervised manner using the DNS Challenge 2020 dataset \citep{reddy2020interspeech}. TorchAudio-Squim also provides NORESQA-MOS, a learning-based estimate of the subjective mean opinion score (MOS) using non-matching references \citep{manocha22c_interspeech}, denoted by TAS MOS. This metric was trained using the Blizzard
    and Voice Conversion Challenges (BVCC) dataset \citep{cooper2021voices}. We use the implementation of TorchAudio-Squim provided in TorchAudio \citep{yang2022torchaudio,hwang2023torchaudio}.
\end{itemize}

The SI-SDR, PESQ, and STOI measures and their nonintrusive versions provided by TorchAudio-Squim are invariant to a scaling of the speech signal estimate. Conversely, the DNSMOS and NORESQA-MOS metrics are very sensitive to a change in the input signal loudness. This sensitivity would make it difficult to compare different speech enhancement methods without a common normalization procedure. We therefore decided to normalize the speech signal estimates at a common loudness of $-30$~LUFS (Loudness Unit Full Scale) before computing all performance scores, using the Python package \texttt{pyloudnorm} \citep{steinmetz2021pyloudnorm}.

\section{Listening test}
\label{sec:listening_test}

The design of the listening test for the CHiME-7 UDASE task followed the \cite{recommendation2003subjective} methodology \citep{loizou2011speech}. We were inspired by \cite{hu2006subjective, hu2007subjective,naderi21_interspeech} who also evaluated speech enhancement algorithms following this methodology.

\subsection{Specifications}

The listening test was conducted in person at the University of Sheffield (UK) between July 17th and August 9th, 2023. Ethics approval was granted by the University of Sheffield Research Ethics Committee (reference number 052938). It involved 32 participants, with self-reported normal hearing with an average age of 37.1 years old (SD 11.8). The participants were separated into 4 panels of 8 listeners. Each panel was associated with a distinct set of 32 audio samples taken from the \texttt{eval/LT} subset of the segmented CHiME-5 dataset (see Table~\ref{tab:chime-5}), resulting in a total of 128 (32 $\times$ 4) audio samples for the entire listening test. For each audio sample, we had 5 different experimental conditions: 4 speech enhancement systems and the unprocessed noisy speech input. 

\begin{figure}[t]
    \centering
    \includegraphics[width=\linewidth]{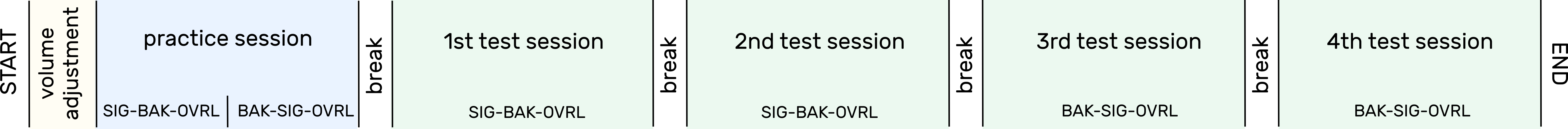}
    \caption{Timeline of the listening experiment. The listening test includes several listening sessions. Each session is made of several trials, and each trial consists of three presentations of the same sound sample. For each presentation, the participant has to give a rating. In this figure, the rating scale order (BAK-SIG-OVRL or SIG-BAK-OVRL) is indicated for each session. The change of the rating scale order in the middle of the session is specific to the practice session.}
    \label{fig:timeline_test}
\end{figure}

\subsection{Methodology and listening experiment}

A complete listening experiment for one participant consisted of 160 trials, where one trial corresponded to a duplet of audio sample and experimental condition (32 audio samples $\times$ 5 experimental conditions = 160 trials). The 160 trials were split into 4 listening sessions of approximately 20 minutes, which correspond to the test sessions in the timeline of Figure~\ref{fig:timeline_test}, and which were separated by short rest periods. 

Each trial consisted of three presentations of the same audio sample, to collect subjective reports on three different rating scales (SIG, BAK, OVRL). Participants were able to listen to the audio sample only once for each presentation in each trial of each session. As shown in Figure~\ref{fig:scales_intructions_test}, in the different presentations participants were instructed to either focus on the speech signal within the audio sample and rate how natural it sounded (SIG rating scale), or focus on the background noise and rate how noticeable or intrusive this background was (BAK rating scale), or attend to both the speech signal and the background noise and rate the overall quality of the audio sample (OVRL rating scale), quality being defined in the perspective of everyday speech communication. The ratings were reported on a 5-point Likert scale. For half of the panels, the order of the presentations was ``SIG, BAK, OVRL" for the first two test sessions, and ``BAK, SIG, OVRL" for the last two. For the other half of the panels, this order was counterbalanced.

A MOS was computed out of 8 votes (one vote per participant) for each triplet \{audio sample, experimental condition, rating scale\}. Overall, this procedure led to 1024 votes (128 audio samples $\times$ 8 votes) for each experimental condition and rating scale.

\subsection{Anchoring phase}
\label{subsec:anchoring}

Before the aforementioned test sessions, the participants performed a practice session, on a material different from the main experiment, which served as an anchoring phase and allowed participants to get familiar with the task. This practice session of 48 trials consisted of the 12 reference conditions described in Table~1 of \citet{naderi21_interspeech} (4 audio samples per reference condition, 2 male and 2 female speakers). The corresponding audio material is available in Microsoft's P.808 Toolkit \citep{naderi2020}. The audio samples for these reference conditions were created by synthetically mixing single-speaker speech and noise signals, following the recommendations of the \citet[Table D.1]{ETSI_TS_103281}. A spectral subtraction algorithm was used to degrade artificially the speech signal at different levels of distortion, while the background noise intrusiveness was controlled by the input SNR. The design of the reference conditions is intended to modulate independently SIG, BAK, and OVRL ratings over the entire five-point range of the Likert rating scales and to equalize the subjective range of quality ratings of all participants.

\subsubsection{Audio presentation and experimental platform}

The participants were seated in a single-walled acoustically-isolated booth. They listened to the audio samples over Sennheiser HD 650 headphones connected to a MOTU M4 audio interface. The loudness of all samples was normalized to -30 LUFS before the listening test, and the listening system was calibrated to a nominal listening level of approximately 78 dBA before each listening session, using a miniDSP EARS headphone test fixture. Before starting the experiment, the participants were able to listen to a few sound samples to verify that the default listening level was comfortable. They were encouraged to not change the default listening level, but they could adjust it with a slider (between -6.0 and +6.0~dB) in case of discomfort. Among the 32 participants, 4 chose to change the default listening level (+0.3, +0.7, +1.1, and -3.0 dB).

We developed an experimental platform based on the JavaScript framework jsPsych \citep{de2023jspsych} to present the audio stimuli and register the participants' ratings in a web browser. This JavaScript experimental platform is provided as an open source library,\footnote{\url{https://github.com/UDASE-CHiME2023/jsPsych-P.835-listening-test} (last accessed: 2024-02-02)} allowing one to reproduce the CHiME-7 UDASE listening experiment using the released audio material. It is also designed to be easily adapted to different audio materials, to help speech enhancement researchers implement ITU-T P.835 listening tests in the future. 

Each presentation of each trial in the experimental platform consisted of three windows: the first window displayed an instruction (e.g., ``Attend ONLY to the SPEECH SIGNAL'') and asked the participant to click on a ``Play sound'' button when ready; the second window displayed a small cross in the middle of the screen while the sound was playing, to let the participant focus on the audio stimulus; the third window (shown in Figure~\ref{fig:scales_intructions_test}) repeated the instruction (e.g., ``Attending ONLY to the SPEECH SIGNAL, select the category which best describes the sample you just heard.'') and presented a 5-position slider to register the vote. The initial position of this slider was randomized for each rating.

\section{Speech enhancement methods}
\label{sec:SE_methods}

This section presents the speech enhancement methods that were submitted to the CHiME-7 UDASE task. In addition to the baseline, we received five submissions from three different teams: The N\&B submission from Northwestern Polytechnical University and ByteDance (China); two submissions, ISDS1 and ISDS2, from Sogang University (Korea); and two submissions, CMGAN-base and CMGAN-FT, from the University of Sheffield (UK).

\subsection{Baseline}

The CHiME-7 UDASE baseline is based on the RemixIT framework of \cite{tzinis2022continual,tzinis2022remixit}. The baseline was developed by first training a supervised teacher model (``OOD teacher") on the out-of-domain LibriMix dataset (see Section~\ref{subsec:librimix}). Then, we fed in-domain CHiME-5 noisy speech recordings to the pre-trained teacher model, to get estimates of the isolated speech and noise signals that will serve as pseudo-labels to train a student model. We synthesized new bootstrapped mixtures by remixing the speech and the permuted noise estimates from the teacher model. Finally, these new mixtures and the corresponding pseudo-labels were used to train a student model for speech enhancement in the target domain, without the need for in-domain reference signals. The teacher and student models are based on the same Sudo-rm-rf sound separation model \citep{tzinis2020sudo,tzinis2022compute}, and they were trained by minimizing the negative SI-SDR loss.

We provided two versions of the student model. The first student model (``RemixIT") was trained using the raw audio segments of the CHiME-5 training set (see Table~\ref{tab:chime-5}). An issue with this approach is that the audio segments do not always contain speech, which may negatively impact the training of the student model. So, we trained a second student model (``RemixIT-VAD") using only the audio segments that were automatically labeled as containing speech by the off-the-shelf voice activity detector Brouhaha \citep{lavechin2022brouhaha}.

Additional information about the baseline is provided in \citet{leglaive2023chime}, and the implementation is available online.\footnote{\url{https://github.com/UDASE-CHiME2023/baseline} (last accessed: 2024-02-02)}

\subsection{NPU and ByteDance submission (N\&B)}

The N\&B submission of \citet{zhang2023nwpu} uses a self-supervised learning approach based on RemixIT \citep{tzinis2022remixit}. Teacher and student models use the Uformer architecture \citep{fu2022uformer}. MetricGAN+ \citep{fu2021metricgan+} is used to mimic the behavior of PESQ or DNSMOS. This GAN is applied to the speech outputs of the student model in RemixIT to ensure enhanced speech with good perceptual quality, where this enhanced speech is used as pseudo-labels in RemixIT. In addition, an unsupervised noise adaptation strategy with data simulation called UNA-GAN \citep{chen2023unsupervised} is used to generate noisy speech in the target domain. Finally, perceptual contrast stretching \citep{chao2022perceptual} is applied to both the input noisy speech during training and the enhanced speech after inference as post-processing. The system is trained on 1-3 speaker mixtures of LibriSpeech utterances plus noise from WHAM! and the VAD-segmented CHiME-5 train set, with about 30\% of the clean speech examples convolved with synthetic room impulse responses. Unlabeled CHiME-5 is used to train the UNA-GAN and the RemixIT student model.
Additional details about the N\&B system can be found in \cite{zhang2023nwpu}.

\subsection{Sogang University submissions (ISDS1 and ISDS2)}

The submission of \citet{jang2023chime} presents two speech enhancement systems, which build on the RemixIT \citep{tzinis2022remixit} pipeline. In the first proposed system, called ISDS1, the U-Net-based network of the Sudo-rm-rf baseline is replaced with a Mossformer architecture \citep{zhao2023mossformer}, having convolution-augmented joint local and global self-attention mechanisms. This architecture can capture the long-range direct interaction between the global intermediate and local features, resulting in a more detailed feature design. In addition to this modification, a speech purification technique is introduced for the self-supervised learning of the student model in RemixIT, leading to the second proposed system, called ISDS2. This is done by predicting the SNR for each frame-level segment, using a pre-trained recurrent neural network, and utilizing these predictions as weights in the training loss of the student model. Additional details about the Sogang systems can be found in \citet{jang2023chime}.

\subsection{University of Sheffield submissions (CMGAN-base and CMGAN-FT)}

Unlike the other submissions, the submission of \cite{close2023university} does not rely on RemixIT. Rather, the method is based on conformer-based metric GAN (CMGAN) \citep{cao2022cmgan}, with two extensions: first, for each epoch, the discriminator was trained on a historical set of past generator outputs; and second, the discriminator was trained to predict the DNSMOS metric score of clean, noisy, and enhanced audio, as well as audio from a pseudo-generator network designed to provide a wider range of metric values. The input to the discriminator is preprocessed by a HuBERT encoder \citep{hsu2021hubert}, and the discriminator also includes losses that measure the mean-squared error between HuBERT representations. The pseudo-generator is only trained with one of the discriminator loss terms, a simple least-squares GAN loss. The base system (``CMGAN-base") is only trained on LibriMix. Another variant, CMGAN fine-tuned (``CMGAN-FT"), was also submitted, which was further fine-tuned on the reverberant LibriCHiME-5 dev set. The provided unlabeled CHiME-5 data was not used during training.
Additional details about the CMGAN systems can be found in \cite{close2023university}.

\begin{table}[!ht]
\centering
\begin{subtable}[t]{\linewidth}
\centering
\begin{tabular}{lccc|cccc}
\cmidrule[1pt]{1-8}
& \multicolumn{3}{c|}{DNSMOS} & \multicolumn{4}{c}{TorchAudio-Squim (TAS)}  \\
& SIG & BAK & OVRL & SI-SDR & PESQ & STOI & MOS \\\cmidrule(lr){2-8}
Input & \cellcolor[HTML]{FEEEDF}3.48 & \cellcolor[HTML]{FFFFFF}2.92 & \cellcolor[HTML]{FFFEFE}2.84 & \cellcolor[HTML]{FFFFFF}-2.1 & \cellcolor[HTML]{FFFFFF}1.31 & \cellcolor[HTML]{FFFFFF}0.65 & \cellcolor[HTML]{FFFFFF}3.04 \\\cmidrule(lr){2-8}
OOD teacher & \cellcolor[HTML]{FFFAF4}3.33 & \cellcolor[HTML]{FCDEC0}3.59 & \cellcolor[HTML]{FFFCF8}2.88 & \cellcolor[HTML]{F9CB9C}\textbf{3.9} & \cellcolor[HTML]{FBD6B0}{\ul 1.56} & \cellcolor[HTML]{F9CB9C}\textbf{0.79} & \cellcolor[HTML]{FAD4AD}3.71 \\
RemixIT & \cellcolor[HTML]{FFFFFF}3.25 & \cellcolor[HTML]{FBDCBB}3.64 & \cellcolor[HTML]{FFFFFF}2.82 & \cellcolor[HTML]{FCDFC1}1.6 & \cellcolor[HTML]{FCE1C5}1.49 & \cellcolor[HTML]{FBDCBD}0.74 & \cellcolor[HTML]{FBDAB9}3.61 \\
RemixIT-VAD & \cellcolor[HTML]{FFFEFC}3.28 & \cellcolor[HTML]{FBDDBD}3.62 & \cellcolor[HTML]{FFFEFD}2.84 & \cellcolor[HTML]{FBDBB9}2.1 & \cellcolor[HTML]{FCE0C4}1.49 & \cellcolor[HTML]{FBDAB8}0.75 & \cellcolor[HTML]{FBDAB8}3.61 \\
CMGAN-base & \cellcolor[HTML]{FBD8B5}{\ul 3.76} & \cellcolor[HTML]{F9CB9C}\textbf{3.97} & \cellcolor[HTML]{FBD7B2}{\ul 3.40} & \cellcolor[HTML]{FEEFE1}-0.3 & \cellcolor[HTML]{FFF8F1}1.35 & \cellcolor[HTML]{FDE8D3}0.71 & \cellcolor[HTML]{FBD6B1}3.68 \\
CMGAN-FT & \cellcolor[HTML]{F9CB9C}\textbf{3.92} & \cellcolor[HTML]{FACEA1}{\ul 3.93} & \cellcolor[HTML]{F9CB9C}\textbf{3.56} & \cellcolor[HTML]{FEF3E8}-0.7 & \cellcolor[HTML]{FFFEFD}1.32 & \cellcolor[HTML]{FDE9D5}0.71 & \cellcolor[HTML]{FAD1A7}{\ul 3.75} \\
ISDS1 & \cellcolor[HTML]{FEF5EC}3.39 & \cellcolor[HTML]{FCDEBF}3.60 & \cellcolor[HTML]{FFFAF6}2.90 & \cellcolor[HTML]{FCDDBE}1.8 & \cellcolor[HTML]{FBDCBC}1.52 & \cellcolor[HTML]{FBDBBA}0.75 & \cellcolor[HTML]{FBDAB7}3.62 \\
ISDS2 & \cellcolor[HTML]{FFFAF6}3.32 & \cellcolor[HTML]{FBD9B6}3.70 & \cellcolor[HTML]{FFFCF8}2.88 & \cellcolor[HTML]{FDEDDB}0.1 & \cellcolor[HTML]{FEF1E4}1.40 & \cellcolor[HTML]{FDE6CF}0.72 & \cellcolor[HTML]{FBD7B2}3.66 \\
N\&B & \cellcolor[HTML]{FEF5EC}3.39 & \cellcolor[HTML]{FACEA1}{\ul 3.93} & \cellcolor[HTML]{FDEEDE}3.07 & \cellcolor[HTML]{FACFA3}{\ul 3.5} & \cellcolor[HTML]{F9CB9C}\textbf{1.62} & \cellcolor[HTML]{FAD4AD}{\ul 0.77} & \cellcolor[HTML]{F9CB9C}\textbf{3.84} \\
\cmidrule[1pt]{1-8}
\end{tabular}
\caption{Results on the single-speaker subset of the CHiME-5 evaluation set (\texttt{eval/1}).}
\label{tab:objective_eval_chime5}
\hspace{1cm}
\end{subtable}

\begin{subtable}[t]{\linewidth}
\centering
\begin{tabular}{lccc|ccc|cccc}
\cmidrule[1pt]{1-11}
& \multirow{2}{*}{SI-SDR} & \multirow{2}{*}{PESQ} & \multirow{2}{*}{STOI} & \multicolumn{3}{c|}{DNSMOS} & \multicolumn{4}{c}{TorchAudio-Squim (TAS)}  \\
& & & & SIG & BAK & OVRL & SI-SDR & PESQ & STOI & MOS \\\cmidrule(lr){2-11}
Input & \cellcolor[HTML]{FEF4E9}6.6 & \cellcolor[HTML]{FFFFFF}1.55 & \cellcolor[HTML]{FDE9D5}0.71 & \cellcolor[HTML]{FDECDA}3.50 & \cellcolor[HTML]{FFFFFF}2.93 & \cellcolor[HTML]{FFF8F1}2.85 & \cellcolor[HTML]{FFFFFF}-5.4 & \cellcolor[HTML]{FFFFFF}1.18 & \cellcolor[HTML]{FFFFFF}0.55 & \cellcolor[HTML]{FFFFFF}3.31 \\\cmidrule(lr){2-11}
OOD teacher & \cellcolor[HTML]{FDECDA}7.8 & \cellcolor[HTML]{FFFEFD}1.57 & \cellcolor[HTML]{FFFDFA}0.65 & \cellcolor[HTML]{FFFAF6}3.30 & \cellcolor[HTML]{FCE0C4}3.60 & \cellcolor[HTML]{FFFCF9}2.79 & \cellcolor[HTML]{F9CB9C}\textbf{-1.2} & \cellcolor[HTML]{FDE8D3}{\ul 1.23} & \cellcolor[HTML]{F9CB9C}\textbf{0.67} & \cellcolor[HTML]{FAD2A9}3.67 \\
RemixIT & \cellcolor[HTML]{FCE2C7}9.4 & \cellcolor[HTML]{FFF8F0}1.68 & \cellcolor[HTML]{FEF5EB}0.67 & \cellcolor[HTML]{FFFFFF}3.23 & \cellcolor[HTML]{FBDCBB}3.69 & \cellcolor[HTML]{FFFFFF}2.74 & \cellcolor[HTML]{FDECDB}-3.8 & \cellcolor[HTML]{FEF3E8}1.20 & \cellcolor[HTML]{FCDFC2}0.63 & \cellcolor[HTML]{FCE3C9}3.54 \\
RemixIT-VAD & \cellcolor[HTML]{FCDEBF}10.1 & \cellcolor[HTML]{FEF7EF}1.69 & \cellcolor[HTML]{FEF3E7}0.68 & \cellcolor[HTML]{FFFCF9}3.27 & \cellcolor[HTML]{FBDCBB}3.69 & \cellcolor[HTML]{FFFCFA}2.79 & \cellcolor[HTML]{FCE3CA}{\ul -3.1} & \cellcolor[HTML]{FEF4E9}1.20 & \cellcolor[HTML]{FBDCBC}0.64 & \cellcolor[HTML]{FCDFC3}3.56 \\
CMGAN-base & \cellcolor[HTML]{FDECDA}7.8 & \cellcolor[HTML]{FFF8F2}1.66 & \cellcolor[HTML]{FDEDDD}0.69 & \cellcolor[HTML]{FBD8B3}{\ul 3.79} & \cellcolor[HTML]{FACEA1}{\ul 3.99} & \cellcolor[HTML]{FAD3AB}{\ul 3.41} & \cellcolor[HTML]{FEEEDF}-4.0 & \cellcolor[HTML]{FFFBF6}1.19 & \cellcolor[HTML]{FCDFC1}0.63 & \cellcolor[HTML]{FAD0A5}{\ul 3.69} \\
CMGAN-FT & \cellcolor[HTML]{FFFFFF}4.7 & \cellcolor[HTML]{FFFDFB}1.59 & \cellcolor[HTML]{FFFFFF}0.64 & \cellcolor[HTML]{F9CB9C}\textbf{3.96} & \cellcolor[HTML]{FAD1A7}3.92 & \cellcolor[HTML]{F9CB9C}\textbf{3.53} & \cellcolor[HTML]{FDEEDE}-4.0 & \cellcolor[HTML]{FFFFFF}1.18 & \cellcolor[HTML]{FCDFC2}0.63 & \cellcolor[HTML]{F9CB9C}\textbf{3.72} \\
ISDS1 & \cellcolor[HTML]{FACFA3}{\ul 12.4} & \cellcolor[HTML]{FDE7D1}{\ul 1.95} & \cellcolor[HTML]{FCE3C9}{\ul 0.73} & \cellcolor[HTML]{FEF7EF}3.35 & \cellcolor[HTML]{FCDEC0}3.64 & \cellcolor[HTML]{FFFBF7}2.81 & \cellcolor[HTML]{FEF3E7}-4.4 & \cellcolor[HTML]{FEF6EE}1.20 & \cellcolor[HTML]{FDE7D1}0.61 & \cellcolor[HTML]{FCDEC0}3.57 \\
ISDS2 & \cellcolor[HTML]{FACFA3}{\ul 12.4} & \cellcolor[HTML]{FEF5EB}1.72 & \cellcolor[HTML]{FDE6CF}0.72 & \cellcolor[HTML]{FFFAF5}3.30 & \cellcolor[HTML]{FBDBBA}3.71 & \cellcolor[HTML]{FFFDFA}2.78 & \cellcolor[HTML]{FFFDFA}-5.2 & \cellcolor[HTML]{FFFEFD}1.18 & \cellcolor[HTML]{FDEBD8}0.60 & \cellcolor[HTML]{FBD7B2}3.63 \\
N\&B & \cellcolor[HTML]{F9CB9C}\textbf{13.0} & \cellcolor[HTML]{F9CB9C}\textbf{2.40} & \cellcolor[HTML]{F9CB9C}\textbf{0.80} & \cellcolor[HTML]{FEF0E2}3.45 & \cellcolor[HTML]{F9CB9C}\textbf{4.03} & \cellcolor[HTML]{FEF0E2}2.98 & \cellcolor[HTML]{FEEEDF}-4.0 & \cellcolor[HTML]{F9CB9C}\textbf{1.29} & \cellcolor[HTML]{FAD0A4}{\ul 0.66} & \cellcolor[HTML]{FAD4AD}3.65 \\\cmidrule(lr){2-11}
Random & -54.8 & 1.14 & 0.00 & 2.86 & 1.08 & 2.22 & -14.4 & 1.21 & 0.49 & 3.15 \\
Oracle & 43.6 & 4.64 & 1.00 & 3.41 & 3.67 & 2.91 & -3.3 & 1.23 & 0.63 & 3.98 \\
\cmidrule[1pt]{1-11}
\end{tabular}
\caption{Results on the reverberant LibriCHiME-5 evaluation set. The DNSMOS and TorchAudio-Squim metrics are computed only on the \texttt{eval/1} subset to be consistent with their single-speaker training condition. The `Random' entry corresponds to the case where white Gaussian noise is taken as the speech estimate, and the `Oracle' entry corresponds to the case where the ground-truth clean speech signal is taken as the estimate. The `Oracle' and `Random' conditions are excluded when defining the color scale and the best and second-best scores.}
\label{tab:objective_eval_librichime5}
\end{subtable}
\caption{\normalsize Mean objective evaluation results on the CHiME-5 (\subref{tab:objective_eval_chime5}) and reverberant LibriCHiME-5 (\subref{tab:objective_eval_librichime5}) datasets. The orange color scale is defined column-wise, the darker the higher. The best scores for each metric are in bold, and second best scores are underlined.}
\label{tab:table1}
\end{table}

\section{Results and discussion}
\label{sec:results}

This section presents and analyzes the objective and subjective evaluation results.

\subsection{Objective evaluation}
\label{subsec:obj_eval}

The mean objective evaluation results are provided in Tables~\ref{tab:objective_eval_chime5} and \ref{tab:objective_eval_librichime5} for the CHiME-5 and the reverberant LibriCHiME-5 datasets, respectively. To be consistent with their training configuration, the supervised DNSMOS and TorchAudio-Squim metrics are computed only on the single-speaker subsets (\texttt{eval/1}). The `Input' condition is obtained by taking the noisy speech signal as the clean speech estimate.

\subsubsection{Analysis of the results according to the metrics}

A first striking observation is the strong dependence of method rankings on the chosen metric. In terms of DNSMOS metrics, the CMGAN submissions perform the best, followed by the N\&B submission. Notably, the N\&B submission excels in SI-SDR, PESQ, and STOI metrics on the reverberant LibriCHiME-5. Meanwhile, the OOD teacher model leads in the TAS SI-SDR, TAS PESQ, and TAS STOI metrics, with the N\&B system following closely. Finally, depending on the dataset, either the CMGAN or N\&B submissions perform best in terms of TAS MOS.

The variability of the results according to the chosen metric makes their interpretation difficult. Therefore, to assess the reliability of the different metrics, we added the `Random' and `Oracle' entries in Table~\ref{tab:objective_eval_librichime5}. These conditions are obtained by taking random white Gaussian noise and the ground-truth reference signal as the clean speech estimate, respectively. As expected, it can be seen that the widely-used intrusive SI-SDR, PESQ, and STOI metrics obtain their minimum and maximum values for the `Random' and `Oracle' conditions, respectively.\footnote{The SI-SDR is theoretically unbounded but its implementation uses machine epsilon to avoid NaN values.} In terms of DNSMOS and TorchAudio-Squim metrics, the `Oracle' condition never obtains the best scores, which might seem surprising at first sight. However, this could be explained by the fact that the quality of the LibriSpeech recordings used to create the reverberant LibriCHiME-5 dataset is not always good. Even if these speech recordings were automatically labeled as `clean' (see the procedure in \cite{panayotov2015librispeech}), they were made by volunteers without any guarantee of the audio quality. 

\paragraph{Analysis of DNSMOS} The `Input' and `Oracle' DNSMOS SIG scores should ideally be the same, as these two conditions include the same speech signals, with or without additive background noise. In practice, this is not the case, but the two values remain close with only a 0.09-point difference. As expected, the `Random' BAK score of 1.08 is by far the lowest among all conditions. However, it is suspicious that the `Random' SIG and OVRL scores are as high as 2.86 and 2.22, considering that these were obtained from white Gaussian noise signals. This behavior could be attributed to the supervised training of the DNSMOS model, which might provide unreliable results on examples that strongly differ from the training conditions.

\begin{figure}[t]
    \centering
    \begin{subfigure}{.49\linewidth}
        \includegraphics[width=\textwidth]{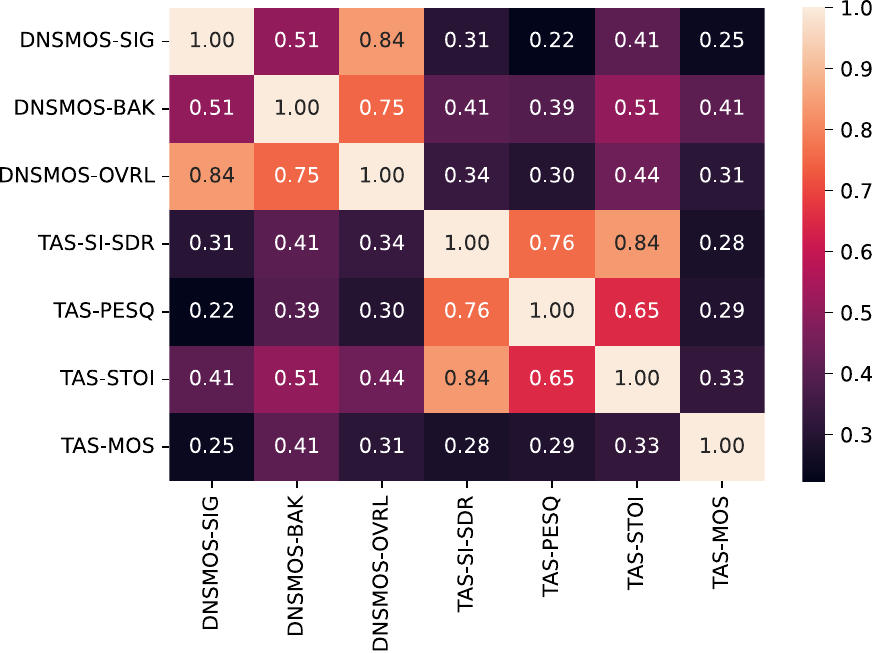}
        \caption{Results on the \texttt{eval/1} subset of the CHiME-5\\ dataset.}
        \label{fig:corr_CHiME-5}
    \end{subfigure} \hfill
    \begin{subfigure}{.48\linewidth}
	    \includegraphics[width=\textwidth]{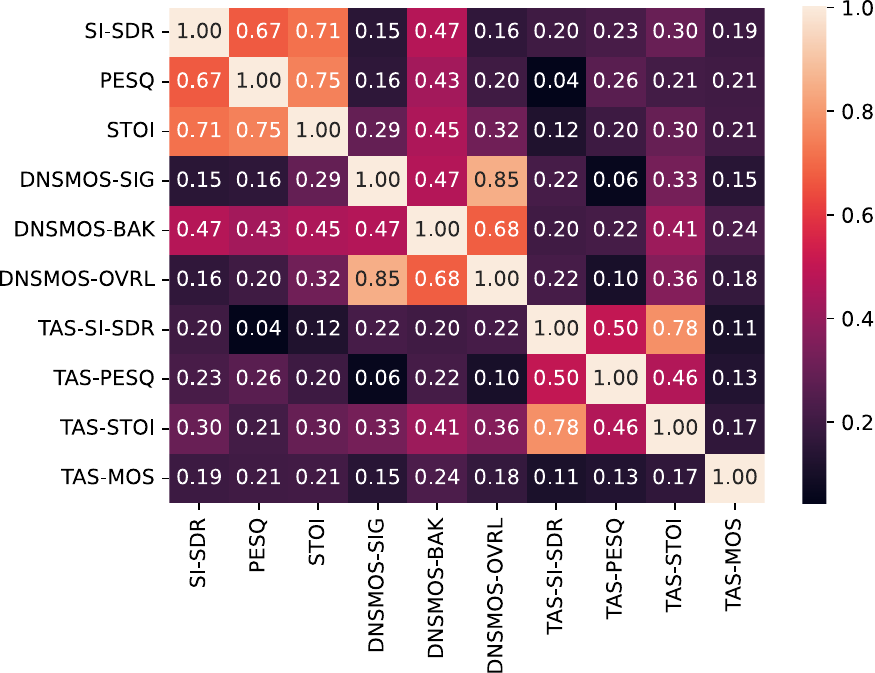}
        \caption{Results on the \texttt{eval/1} subset of the reverberant LibriCHiME-5 dataset.}
	    \label{fig:corr_libriCHiME-5}
	\end{subfigure}
    \caption{Pairwise Pearson correlation coefficient of the objective performance metrics.}
    \label{fig:corr_objective_metrics}
\end{figure}

\paragraph{Analysis of TorchAudio-Squim} There is a significant inconsistency between the results obtained with the SI-SDR, PESQ, and STOI metrics and their nonintrusive equivalents as provided by TorchAudio-Squim. For instance, on reverberant LibriCHiME-5, the N\&B system obtains the best performance in terms of SI-SDR but one of the worst performances in terms of TAS SI-SDR.\footnote{The N\&B system was ranked first with the listening test, as will be discussed later.} To investigate this inconsistency, we computed the Pearson correlation coefficient (PCC) between each pair of metrics for each dataset (we only used the single-speaker evaluation subset for reverberant LibriCHiME-5). The PCC values are shown in Figures~\ref{fig:corr_CHiME-5} and \ref{fig:corr_libriCHiME-5} for the CHiME-5 and reverberant LibriCHiME-5 datasets, respectively. On the latter, the PCC of the TorchAudio-Squim metrics and their intrusive equivalents ranges from 0.20 to 0.30, which is much lower than the 0.95 to 0.98 values reported in \citet{kumar2023torchaudio}. This very low correlation suggests that the TorchAudio-Squim SI-SDR, PESQ, and STOI estimates are not sufficiently reliable on the CHiME-7 UDASE data. These metrics are obtained from a supervised neural model trained on the DNS Challenge 2020 dataset, which may not generalize well to the CHiME-7 UDASE data. Interestingly, the TAS MOS seems to be more reliable. Indeed, the `Random' condition score (3.15) is probably too high but it is also the worst among all conditions, the second worst score corresponds to the `Input' condition (3.31), and the best score is obtained with the `Oracle' condition (3.98). As can also be seen in Figure~\ref{fig:corr_objective_metrics}, TAS MOS shows a limited correlation with TAS SI-SDR, TAS PESQ, and TAS STOI, with a PCC ranging from 0.11 to 0.33. In contrast, these three latter metrics exhibit a moderate-to-high correlation with each other, with PCC values between 0.50 and 0.84.

\subsubsection{Analysis of the results according to the datasets}

As discussed above, the ranking of the methods varies a lot according to the chosen metric. However, this ranking seems to be quite similar across the two datasets for a given metric, as can be seen by comparing the color scales in Tables~\ref{tab:objective_eval_chime5} and \ref{tab:objective_eval_librichime5}. In particular, the DNSMOS scores on both datasets are very close for all conditions. For instance, the `Input' DNSMOS scores are (SIG, BAK, OVRL) = (3.48, 2.92, 2.84) on CHiME-5, and (SIG, BAK, OVRL) = (3.50, 2.93, 2.85) on reverberant LibriCHiME-5. This similarity is less obvious from the TorchAudio-Squim metrics, but as discussed above these seem to be less reliable for our specific use case. It can also be seen that the best-performing systems for each metric are globally the same on the CHiME-5 and reverberant LibriCHiME-5. These similarities of the results on the two evaluation datasets suggest that the reverberant LibriCHiME-5 dataset is sufficiently close to the target domain as defined by the CHiME-5 recordings. 

\subsubsection{Summary}

Overall, we can conclude from the objective evaluation that supervised nonintrusive metrics should be used with caution, because of potential generalization issues. In our specific context, this seems to be particularly true for the nonintrusive SI-SDR, PESQ, and STOI estimates of TorchAudio-Squim. Based on the observations of the previous paragraph, we believe it is reasonable to rely on intrusive metrics computed on the reverberant LibriCHiME-5 dataset to approximate the in-domain performance of speech enhancement methods in the context of the CHiME-7 UDASE task. 

In summary, the objective evaluation reveals that the CMGAN entries yield the best results in terms of DNSMOS metrics on both datasets. Moreover, on the reverberant LibriCHiME-5, the N\&B submission leads in SI-SDR, PESQ, and STOI metrics, followed by the ISDS1 system. Regarding the TAS MOS metric, N\&B outperforms others on the CHiME-5 dataset, while CMGAN-FT takes the lead on the reverberant LibriCHiME-5. A detailed discussion of the baseline results is available in \citet{leglaive2023chime}. The disagreement between the metrics shows the importance of performing a subjective evaluation, whose results will be described in the next section.

\subsection{Subjective evaluation}

\begin{figure*}[t]
    \includegraphics[width=\linewidth]{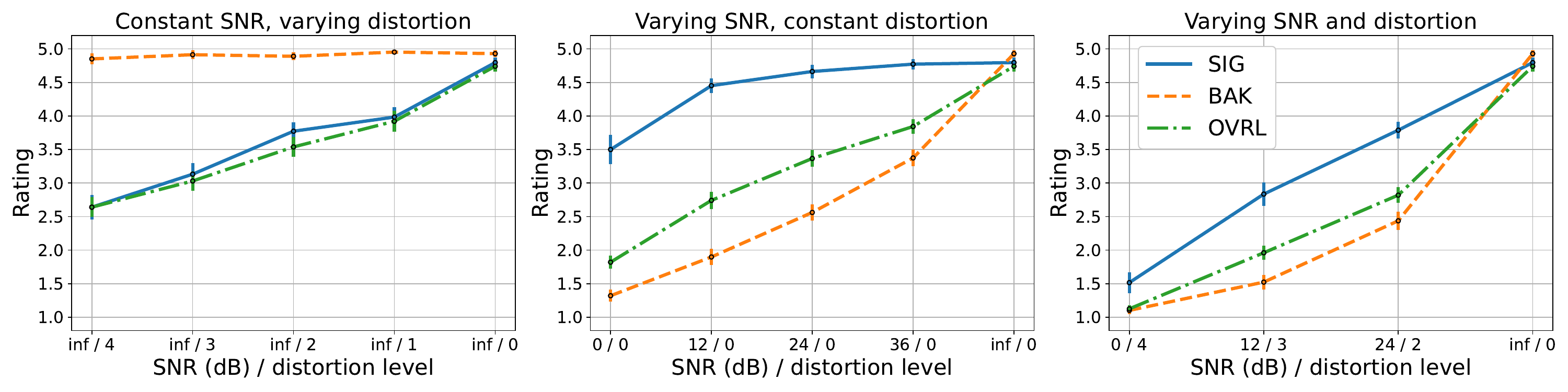}
    \caption{Mean results (dots) and 95\% confidence intervals (bars) for the anchoring phase.}
    \label{fig:achoring_phase}
\end{figure*}

\subsubsection{Anchoring phase}

As introduced in Section~\ref{subsec:anchoring}, during the anchoring phase of the listening test, all participants listened to the same stimuli. These stimuli had been synthetically created by varying the SNR only (0, 12, 24, 36 dB or infinity), the speech distortion level only (between 4 for the highest level of distortion and 0 for no distortion), or both. This led to the 12 conditions described in Table 1 of \citet{naderi21_interspeech}. For each condition, we have 4 stimuli (2 male and 2 female speakers). These 12 conditions are grouped into three sets: (i) Varying SNR, constant distortion level (no distortion); (ii) Constant SNR (infinite, i.e., noise-free), varying distortion level; (iii) Varying SNR, varying distortion level.

The results of this anchoring phase averaged over the 32 participants are shown in Figure~\ref{fig:achoring_phase} for the three different sets of conditions. The results are very similar to those of \citet[Figure~2]{naderi21_interspeech}, which were themselves shown to be highly correlated with the ones reported in  \citet{S4-150762}. When the SNR is fixed and the distortion level varies (left figure), it can be seen that the BAK curve is very flat, as expected. However, the SIG and OVRL ratings do not vary over the entire five-point range of the rating scale. When the distortion level is fixed and the SNR varies (middle figure), we can see that the SIG ratings also vary, which should not be the case. This indicates that some participants tend to confound noise intrusiveness with speech signal quality, especially at the lowest SNRs (0 and 12 dB). When both SNR and distortion levels vary (right figure), the SIG, BAK, and OVRL ratings vary over the entire five-point range, as expected.

Among the 32 participants of the listening test, 4 were older than 50. As hearing thresholds can deteriorate significantly after this age, we verified their rating patterns obtained during the anchoring phase by visually comparing them with those of other younger subjects. We did not find any obvious rating bias for these older subjects.

\subsubsection{Test conditions}

Four speech enhancement methods were selected as experimental conditions for the listening test described in Section~\ref{sec:listening_test}: CMGAN-FT, N\&B, ISDS1, and RemixIT-VAD. The selection procedure described in the rules of the CHiME-7 UDASE task was the following:\footnote{PESQ, STOI, and TorchAudio-Squim metrics were not considered originally for the evaluation of the challenge submissions.}
\begin{itemize}[]
\item We selected the top 3 entries in terms of DNSMOS OVRL score on CHiME-5. In case of multiple entries for the same team, we only kept the best one. This led to S1 = \{CMGAN-FT, N\&B, ISDS1\}.
\item We selected the top 3 entries in terms of SI-SDR on reverberant LibriCHiME-5. In case of multiple entries for the same team, we only kept the best one. This led to S2 = \{N\&B, ISDS1, RemixIT-VAD\}.
\item The union of S1 and S2 gave the above-listed systems selected for the listening test.
\end{itemize}
In addition to these four speech enhancement methods, the listening test included a 5th test condition corresponding to the input unprocessed noisy speech.

\subsubsection{Results}

The subjective BAK MOS, SIG MOS, and OVRL MOS results are shown in Figures~\ref{fig:listening_test_BAK}, \ref{fig:listening_test_SIG}, and \ref{fig:listening_test_OVRL}, respectively. The results are shown with boxplots, violin plots, and mean values. In each figure, the systems are ranked according to their mean results, which are computed from 128 MOS for each system, each MOS being computed from 8 votes. 

\subsubsection{Statistical analysis and discussion}

Repeated measures analyses of variance (ANOVA) were conducted on each rating scale separately, to assess the effect of the experimental conditions (four different systems plus the original audio input as baseline) on the subjective rating judgments. Mauchly's test of sphericity was applied and a Greenhouse-Geisser correction was applied when appropriate. 
The alpha level to accept significance was set at $\alpha = 0.01$. 
\revision{This stringent level was chosen to reduce the risk of Type 1 errors, that is, reporting differences between algorithms when there were none \citep{maier2022justify}.  Reporting follows the \citet{APA2022} guidelines, with all $p$ values less than $0.001$ reported as $p < 0.001$.}
Effect sizes are reported with the $\eta^2$ statistics. When a significant main effect was found, post-hoc comparisons were conducted to compare the ratings obtained by the different experimental conditions, with a Holm-Bonferroni correction.

\begin{figure*}[t]
    \centering
    \begin{subfigure}{.8\linewidth}
        \includegraphics[width=\textwidth]{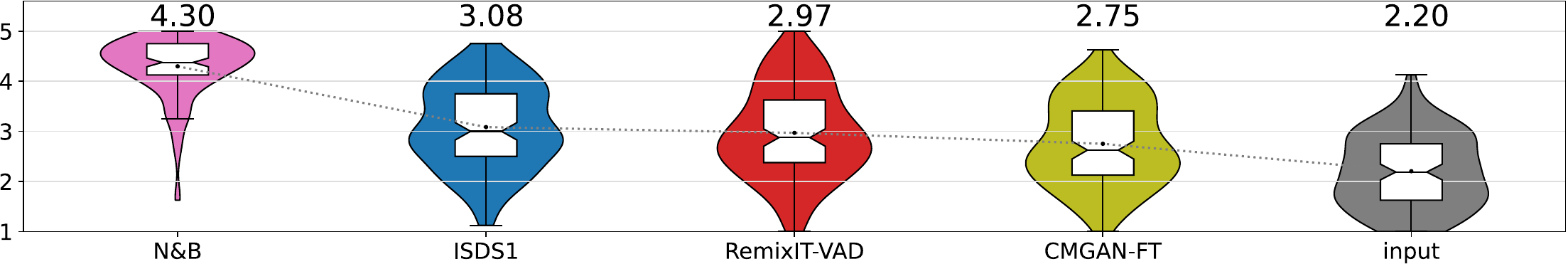}
        \caption{BAK MOS results.}
        \label{fig:listening_test_BAK}
    \end{subfigure}
    \\[.5cm]
    \begin{subfigure}{.8\linewidth}
        \includegraphics[width=\textwidth]{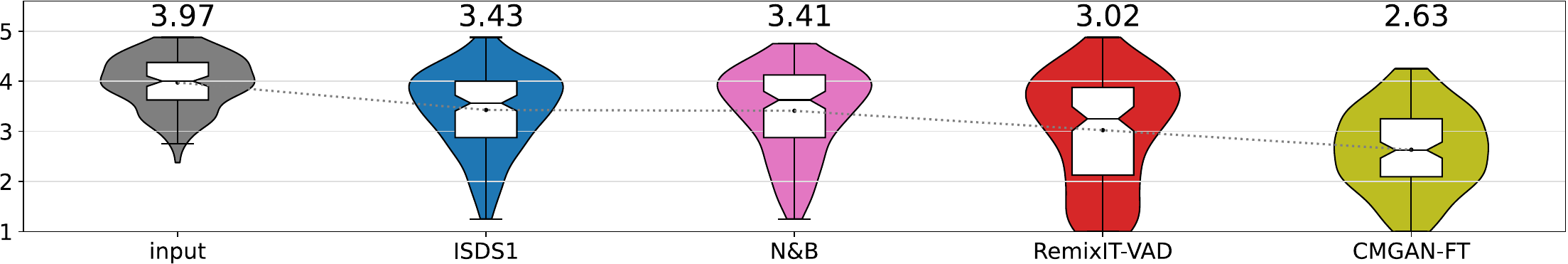}
        \caption{SIG MOS results.}
        \label{fig:listening_test_SIG}
    \end{subfigure}
    \\[.5cm]
    \begin{subfigure}{.8\linewidth}
        \includegraphics[width=\textwidth]{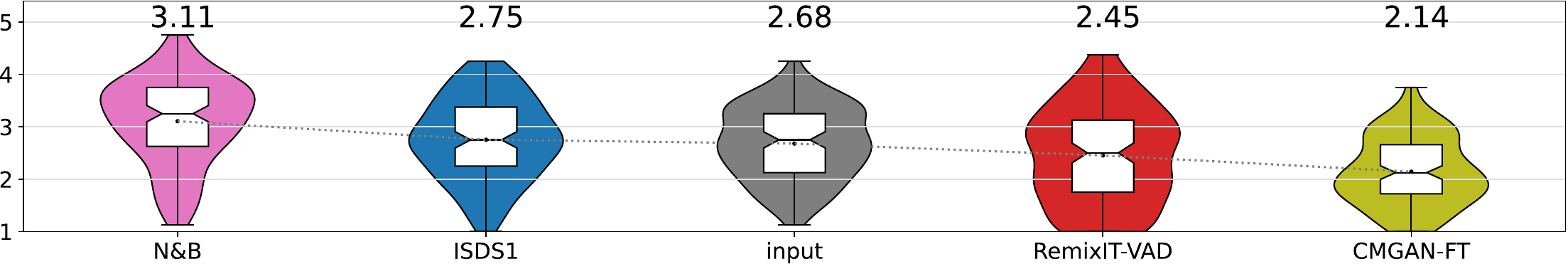}
        \caption{OVRL MOS results.}
        \label{fig:listening_test_OVRL}
    \end{subfigure}
    \caption{Boxplots, violin plots, and mean results for the subjective BAK (\subref{fig:listening_test_BAK}), SIG (\subref{fig:listening_test_SIG}), and OVRL (\subref{fig:listening_test_OVRL}) mean opinion scores of the ITU-T P.835 listening test. Black dots and numbers above the box/violin plots correspond to the mean results. In each figure, the systems are ranked according to their mean results.}
    \label{fig:listening_test_results}
\end{figure*}

For background noise reduction ratings, as measured by the BAK MOS, there was a highly significant effect of condition ($F(2.68,31) = 342.18, p < 0.001, \eta^2 = 0.92$). Furthermore, the post-hoc comparisons showed that all systems were effective in reducing noise intrusiveness compared to the baseline input condition (all $p < 0.001$). Moreover, all systems performed differently from each other, except RemixIT-VAD and ISDS1 which performed the same ($p=0.052$). This is consistent with the fact that notches for the boxplots of these two conditions overlap in Figure~\ref{fig:listening_test_BAK}. The noise reduction performance of the N\&B system is very high, on average more than 1 point above the second-best system ISDS1. As can be seen from the violin plot corresponding to the N\&B system in Figure~\ref{fig:listening_test_BAK}, the BAK MOS distribution is strongly compressed toward the maximum value of 5.

Regarding the speech signal quality, as measured by the SIG MOS, there was again a highly significant effect of condition ($F(2.4,31) = 117.31, p < 0.001, \eta^2 = 0.79$). The post-hoc comparisons showed that all systems significantly degraded the speech quality compared to the unprocessed input signals (all $p < 0.001$). This was expected because even though the original input condition included noise, it was not processed in any way that could induce speech distortion. So, the ideal outcome for the SIG MOS metric for the various systems would be to remain as close as possible to the ratings of the original input condition. This ideal outcome was never reached. The best-performing systems on this metric were ISDS1 and N\&B (no difference between them $p=0.800$), followed by RemixIT-VAD and CMGAN-FT (all $p < 0.001$, see Figure~\ref{fig:listening_test_SIG}).

Finally, for the overall quality ratings, as measured by the OVRL MOS, there was also a highly significant effect of condition ($F(2.26,31) = 79.03, p < 0.001, \eta^2 = 0.72$). We can see in Figure~\ref{fig:listening_test_OVRL} that the original audio input condition, which was ranked first in terms of SIG MOS and last in terms of BAK MOS, is now in the middle of the overall quality ranking. This confirms that the overall perceptual quality is a compromise between the distortion of the speech signal and the suppression of the noise. All systems were found to perform significantly differently in terms of overall quality (all $p < 0.001$). However, only the N\&B system significantly improved the overall quality compared to the original audio input condition ($ p < 0.001$). There was no difference observed between the ISDS1 system ratings and the original audio input ratings ($p=0.20$), while the remaining two systems degraded the overall quality compared to the original audio input ($ p < 0.001$). N\&B and ISDS1 have almost identical distortion as measured by the SIG MOS (Figure~\ref{fig:listening_test_SIG}), so the better overall performance for N\&B is likely due to much better noise suppression, as measured by the BAK MOS (Figure~\ref{fig:listening_test_BAK}). The CMGAN-FT and RemixIT-VAD systems were judged to degrade the overall quality, probably because both their noise suppression and their speech signal quality were not sufficiently good. In summary, among the four speech enhancement methods, only the N\&B system succeeded in significantly improving the overall quality compared to the unprocessed noisy speech.

\subsubsection{Comparison with the objective evaluation}

\begin{figure}[t]
	\centering
    \includegraphics[width=.5\linewidth]{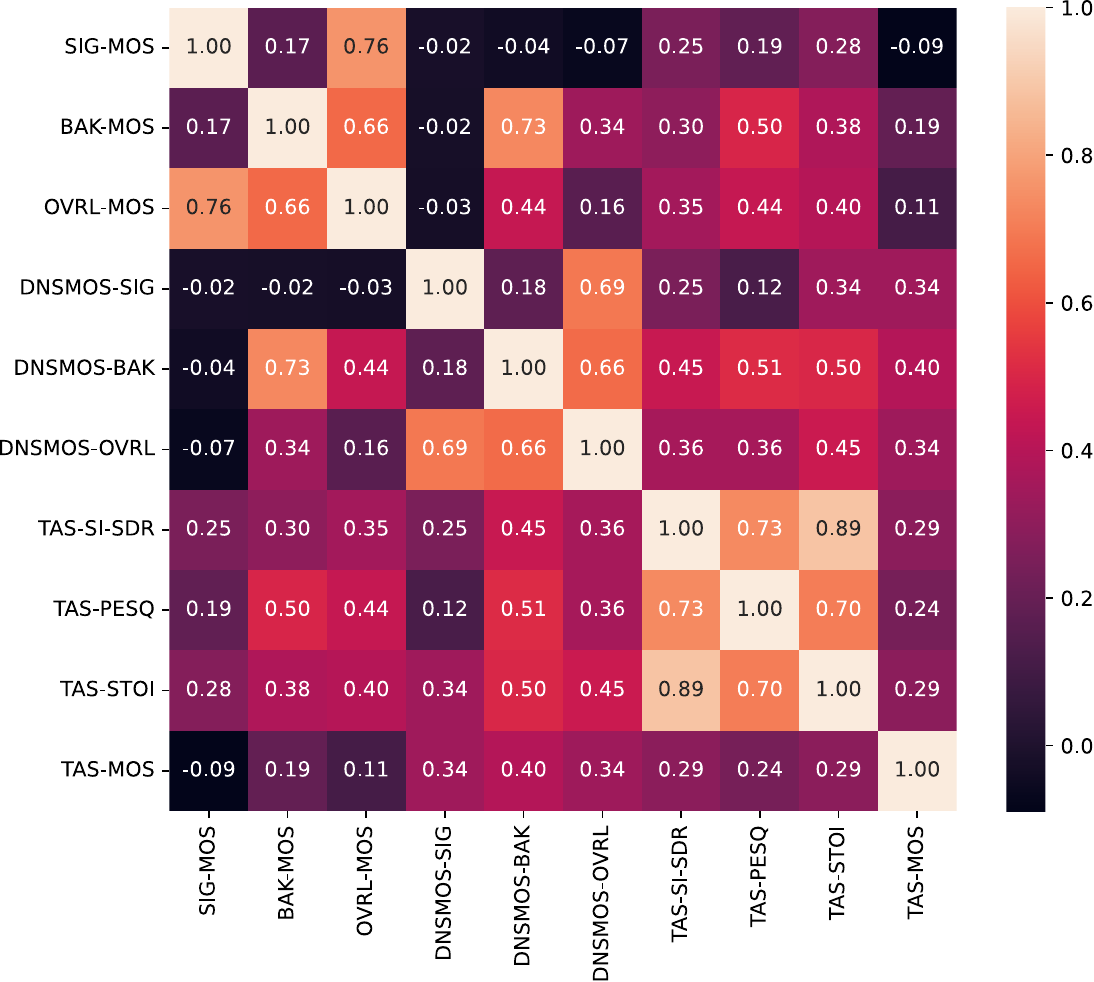}
    \caption{Pearson correlation between objective and subjective evaluation results, computed using only single-speaker samples (115 samples in total).}
    \label{fig:corr_listening_test}
\end{figure}

It can be interesting to compare the results of the objective and subjective evaluations. The CMGAN submissions, which were found to perform best in terms of DNSMOS metrics on both CHiME-5 and reverberant LibriCHiME-5 datasets, are the worst-performing systems according to the listening test (see Figure~\ref{fig:listening_test_results}). As detailed in~\cite{close2023university}, this is probably due to the training of the CMGAN models that rely on the optimization of the supervised DNSMOS metrics. It appears that the models learned to maximize the metrics while providing output speech signals of unsatisfactory quality. This is probably because the DNSMOS model is supervised and it can thus provide unexpected or incoherent results when fed with signals that are too far from its training conditions. On the contrary, the objective evaluation results using standard objective performance metrics (SI-SDR, PESQ, and STOI) computed on the close-to-in-domain reverberant LibriCHiME-5 dataset are consistent with the subjective evaluation results on the CHiME-5 dataset, in terms of ranking of the methods. This again confirms that reverberant LibriCHiME-5 can be used to approximate in-domain performance in the context of the CHiME-7 UDASE task.

Finally, Figure~\ref{fig:corr_listening_test} shows the PCC between the subjective SIG, BAK, OVRL MOS, and the nonintrusive DNSMOS and TorchAudio-Squim metrics. It can be seen that, apart from a PCC of 0.73 between the DNSMOS BAK score and the subjective BAK MOS, the nonintrusive objective metrics poorly correlate with the subjective ones (PCC lower than or equal to 0.50). The DNSMOS BAK score was also found to be the most reliable nonintrusive metric when analyzing the objective evaluation results in Section~\ref{subsec:obj_eval}. Interestingly, the correlation between the subjective SIG, BAK, and OVRL metrics (0.17, 0.76, 0.66) is very close to the correlation between the equivalent objective metrics as provided by DNSMOS (0.18, 0.69, 0.66). This suggests that DNSMOS has appropriately learned the correlation between the SIG, BAK, and OVRL scores, but unfortunately, this is not sufficient to ensure good generalization, in particular on the CHiME-7 UDASE data.

\section{Conclusion}
\label{sec:conclusion}

Fully-supervised speech enhancement models are trained -- and most of the time also evaluated -- using only synthetic data, which cannot always capture the diversity of real-world speech recordings. A strong mismatch between the synthetic training domain and the real test domain will significantly affect the performance of the model. To address this issue, the CHiME-7 UDASE task aimed to leverage real-world unlabeled noisy speech recordings from the test domain for the unsupervised adaptation of speech enhancement models. Evaluating unsupervised domain adaptation methods for speech enhancement is by definition a challenging task because the ground-truth clean speech signals in the target domain are not available. In this paper, we presented the methodology and analyzed the results of the objective and subjective evaluations conducted in the CHiME-7 UDASE task. 

Objective evaluation in the target domain (defined by the CHiME-5 recordings) relied on the recent DNSMOS P.835 and TorchAudio-Squim nonintrusive performance metrics. Complementarily, we developed the synthetic labeled reverberant LibriCHiME-5 dataset for objective evaluation with common intrusive metrics (SI-SDR, PESQ, STOI) on close-to-in-domain data. The subjective evaluation consisted of an ITU-T P.835 listening test, which was implemented in a JavaScript experimental platform and conducted in person at the University of Sheffield (UK). The subjective evaluation revealed the difficulty of the CHiME-7 UDASE task. Among the four speech enhancement systems that were evaluated in the listening test, only one succeeded in improving the overall quality compared to the unprocessed noisy speech. All systems successfully reduced the background noise but always at the expense of increased distortion, which eventually affected the overall perceived quality.

The analysis of the results revealed that the DNSMOS P.835 and TorchAudio-Squim nonintrusive performance metrics should be used with caution. Their effectiveness was demonstrated for the specific benchmarks presented in \cite{reddy2022dnsmos} and \cite{kumar2023torchaudio}. However, except for the DNSMOS BAK score, these metrics were shown to poorly correlate with the subjective ratings. On the contrary, the ranking of speech enhancement methods on reverberant LibriCHiME-5 using more traditional intrusive objective performance metrics was similar to the ranking based on subjective evaluation. This shows that the reverberant LibriCHiME-5 dataset can be used for in-domain evaluation of speech enhancement models adapted to the unlabeled CHiME-5 dataset. While this is useful practically, it is not entirely satisfying as unsupervised domain adaptation methods should in principle be evaluated on the unlabeled data of the target domain, which is very challenging methodologically \citep{you19tocards,musgrave2021unsupervised}. Further research is needed to address the challenge of training and evaluation of speech enhancement models without clean speech labels.

\appendix

\section{Experimental platform for the listening test}

\begin{figure}[t]
    \centering
    \begin{subfigure}{.85\linewidth}
        \fbox{\includegraphics[width=.95\textwidth]{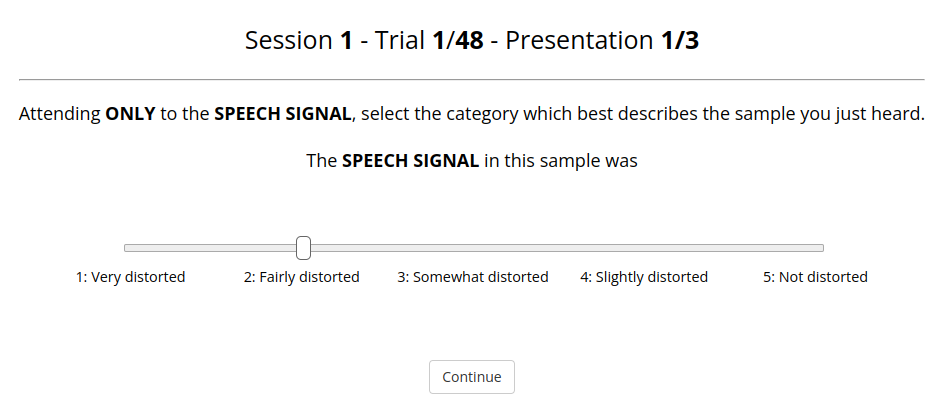}}
        \caption{SIG rating scale and instruction.}
        \label{fig:SIG}
    \end{subfigure}
    \\
    \begin{subfigure}{.85\linewidth}
        \fbox{\includegraphics[width=.95\textwidth]{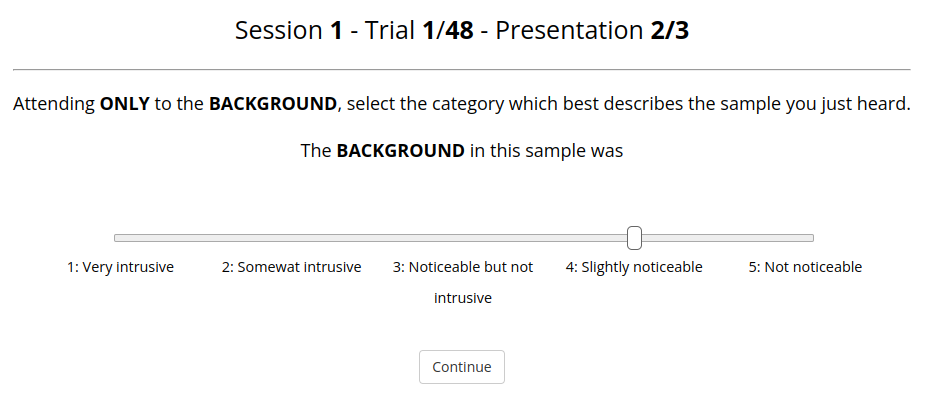}}
        \caption{BAK rating scale and instruction.}
        \label{fig:BAK}
    \end{subfigure}
    \\
    \begin{subfigure}{.855\linewidth}
        \centering
        \fbox{\includegraphics[width=.95\textwidth]{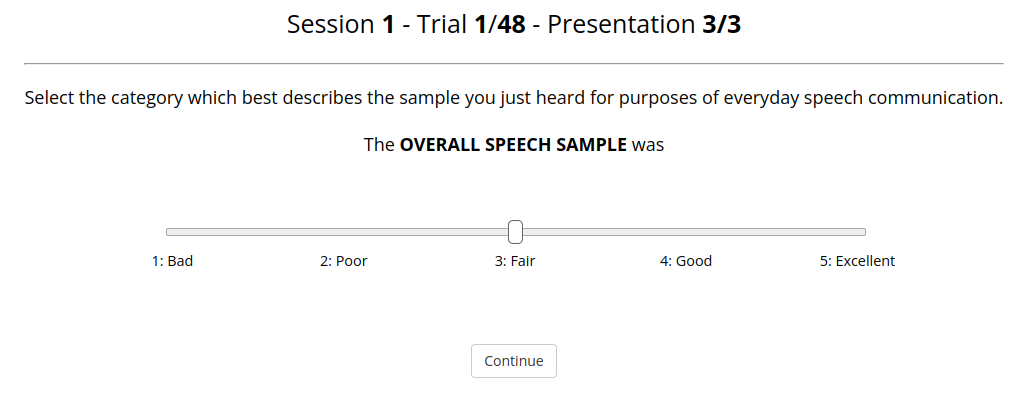}}
        \caption{OVRL rating scale and instruction.}
        \label{fig:OVRL}
    \end{subfigure}
    \revision{\caption{Windows of the experimental platform used for the listening test. Each window presents an instruction and a rating scale following the ITU-T P.835 recommendation, and a 5-position slider to register the vote.}
    \label{fig:scales_intructions_test}
    }
\end{figure}

\revision{Screenshots of the JavaScript experimental platform developed for the listening test of the CHiME-7 UDASE task are shown in Figure~\ref{fig:scales_intructions_test}.}

\section*{Acknowledgments}

\revision{The authors thank the CHiME Steering Group for their support in the organization of the CHiME-7 UDASE task and the participants of the listening test organized at the University of Sheffield. S. Leglaive acknowledges support from the French National Research Agency (ANR) as a part of the DEGREASE project (ANR-23-CE23-0009).}

\bibliography{refs}

\end{document}